\documentclass[aps,pra,twocolumn,showpacs,groupedaddress]{revtex4-1}   % for review and submission with two columns
\usepackage{hyperref}  % needed for live references\
\hypersetup{
colorlinks=true,
urlcolor=blue,
linkcolor=blue, 
pdftoolbar=true, 
filecolor=red, 
citecolor=blue} 
\usepackage{graphicx}  % needed for figures
\usepackage{dcolumn}   % needed for some tables
\usepackage{bm}        % for math
\usepackage{amssymb}   % for math
\usepackage{verbatim}  % for comment
\usepackage{epstopdf}
\usepackage{color}
\newcommand{\ket}[1]{\ensuremath{\left| #1 \right\rangle}}

\newcommand{\beq}{\begin{equation}}
\newcommand{\eeq}{\end{equation}}
\newcommand{\bea}{\begin{eqnarray}}
\newcommand{\eea}{\end{eqnarray}}

\newcommand{\eq}[1]{{(\ref{#1})}}
\newcommand{\commentout}[1]{{}}

\newcommand{\half}{{\hbox{$\frac{1}{2}$}}}

\definecolor{red}{rgb}{1,0,0}

\begin{document}

\title{One- and two-atom states in a rotating ring lattice}

\author{Juha Javanainen}
\author{Otim Odong}
\affiliation{Department of Physics, University of Connecticut, Storrs, Connecticut 06269-3046}

\author{Jerome C. Sanders}
\affiliation{University of Science and Arts of Oklahoma, Chickasha, Oklahoma 73018-5371}

\begin{abstract}
We study the states of one and two atoms in a rotating ring lattice in a Hubbard type tight-binding model. The model is  developed carefully from basic principles in order to properly identify the physical observables. The one-particle ground state may be degenerate and represent a finite flow velocity depending on the parity of the number of lattice sites, the sign of the tunneling matrix element, and the rotation speed of the lattice. Variation of the rotation speed may be used to control one-atom states, and leads to peculiar behaviors such as wildly different phase and group velocities for an atom. Adiabatic variation of the rotation speed of the lattice may also be used to control the state of a two-atom lattice dimer. For instance, at a suitably chosen rotation speed both atoms are confined to the same lattice site.

\end{abstract}

% 03.75.Lm (03.75.-b is Matter waves) Tunneling, Josephson effect, Bose-Einstein condensates in periodic potentials, solitons, vortices, and topological excitations
% 37.10.Jk Atoms in optical lattices
% 05.30.Jp Boson systems
% 05.50.+q Lattice theory and statistics

\maketitle

\section{Introduction}\label{INTRODUCTION}
Optical lattices have enhanced both AMO and condensed matter physics, and will undoubtedly continue to do so for a long time. Theoretical analyses of optical lattices routinely resort to periodic boundary conditions, as if the lattice folded back onto itself into a ring. Usually this is just a matter of convenience, but the topology of a torus may have a profound effect on the physics. As an example, the phase winding of a superfluid around a ring cannot change discontinuously, which ultimately stabilizes a persistent current. Ring traps for atoms have in fact been demonstrated~\cite{RYU07,BRU11,RAM11,SHE11} and employed to study superfluid flow~\cite{RYU07,RAM11}. An intriguing technique whereby a rapidly moving optical trap, when averaged over time, may ``paint''  not only a ring trap but also a virtually arbitrary time dependent structure, in particular, a rotating ring lattice, has also been demonstrated~\cite{HEN09}. A ring lattice with a precisely set and possibly small number of sites rotating at a controllable speed is within the reach of current experimental techniques.

The other background element here is the lattice dimer made of two atoms in an optical lattice. Both the site-to-site tunneling and the atom-atom interactions could be controlled, by adjusting lattice parameters and with the aid of a Feshbach resonance. A lattice dimer may thus make a tailored custom molecule. The experimental demonstration of a dimer bound by repulsive atom-atom interactions~\cite{WIN06} has in part motivated theoretical work from several groups~\cite{ORS05,WIN06,GRU07,PII08,NYG08,NYG08a,VAL08,VAL10,JAV10,SAN11,ODO11}. As usual, we have applied periodic boundary conditions in our analysis~\cite{JAV10,SAN11,ODO11}, which is convenient and permissible when the lattice is long and the boundary conditions cannot matter in practice. However, the boundary conditions are important if the lattice truly  is a finite-size ring, and some of our technical assumptions such as the even number of lattice sites~\cite{JAV10} need to be re-examined. Besides, rotation of the lattice could provide another handle for controlling the molecules.

In the present paper we first, Sec.~\ref{HUBBARD}, study systematically the effects of the rotation of the lattice on a Hubbard type mode by expanding on our earlier coordinate-transformation arguments~\cite{WAN07}. This groundwork allows us to identify the physical observable for one and two atoms in a rotating ring lattice, Secs.~\ref{ONEATOM} and~\ref{TWOATOMS}. Thermal preparation and adiabatic variation of the rotation speed prove to be effective methods to control the state of both one- and two-atom systems, and lead to quite a few perhaps surprising results. For instance, by slowly spinning up the lattice one cannot change the speed with which an atom emerges after it is released from the lattice, but a localized atomic wave packet will {almost} track the varying rotation speed; and at certain rotation speeds a bound dimer of two atoms is confined to a single lattice site. We conclude in Sec.~\ref{CONCLUSION} with a few brief comments.

\section{Hubbard type models}\label{HUBBARD}

Several methods have been used in the past to address the effects of rotation on optical lattices~\cite{WU04,BHA06,WAN07}, and by now the observation that the rotation leads to phase factors in site-to-site tunneling matrix elements has the force of folklore. We find the same basic result, but emphasize that one needs to  keep track of the physical observables of the system carefully.

One may resort to the formal similarity between rotation and magnetic field, say, in that the Coriolis force derives from a vector potential, and argue basically in terms of minimal-coupling substitution~\cite{WU04}; develop for the lattice  an approximation to the usual $-\hbox{\boldmath$\omega$}\cdot{\bf L}$ term that emerges in a transformation to the rotating frame~\cite{BHA06}; or analyze coordinate transformations directly~\cite{WAN07,ZHO06}. All of these methods have their problems, however. For instance, in quantum mechanics coordinate transformations are mathematically unitary transformations and change the appearance of quantum mechanics. Similarly, minimal substitution is carried out in a given fixed gauge, but in quantum mechanics a transformation to a different gauge is also a unitary transformation. The following question exemplifies the core of the problem: In which unitarily transformed version of quantum mechanics is $e^{i{\bf p}\cdot{\bf r}}$ the wave function representing a particle with the velocity ${\bf p}/m$? The debates about the  ${\bf p}\cdot{\bf A}$ and ${\bf d}\cdot{\bf E}$ forms of the dipole interaction that have erupted periodically in the past are a manifestation of this type of ambiguity~\cite{COH92}.  We resolve such issues by carefully expanding on our earlier coordinate-transformation approach~\cite{WAN07}. In the end, we will keep track of three reference frames.

We start from the assumption that the motion of the atoms is confined to a torus with the circumference $L$. In the limit of an asymptotically strong transverse confinement, the coordinate along the direction of the torus, $x$, remains the only relevant degree of freedom. In principle one might consider angular momenta, but for sufficiently tight transverse confinement we may model the motion in the $x$ direction simply as one-dimensional translation. The physics takes place over the interval $[-L/2,L/2)$, but here we imagine that the coordinate $x$ ranges over the entire real axis and impose periodic boundary conditions over the distance $L$ to account for the topology of the torus.  The inner product of one-atom wave functions is defined as
\beq
(\psi,\phi) = \int_{-L/2}^{L/2}dx\,\psi^*(x)\phi(x)\,,
\label{INNERPROD}
\eeq
from which the Hilbert space structure follows.

Suppose next that  we have a stationary potential $V(x)$ along the direction of the torus. We take the potential $V(x)$ to be periodic over the distance $L$ and also over a shorter distance $a=L/N$, so that we have an $N$-site lattice with the lattice spacing $a$ and periodic boundary conditions from end to end of the lattice. We denote the sites of the lattice, e.g., minima of the potential $V(x)$, by $x_n$, and for future reference also define the wave vectors $k_n$. We set
\beq
x_n = n a,\quad
k_n = \frac{2\pi n}{L}\,.\eeq
The choice here (by no means unique) is that, unless otherwise specified,  for an even number of the sites in the lattice  the index $n$ ranges from $-N/2$ to $N/2-1$ in unit steps, while for an odd number of lattice sites $N$ the range is from $-N/2-1/2$ to $N/2-1/2$. The wave vectors thus run over the first Brillouin zone of solid-state physics.

Assume now that the potential $V(x)$ is made to rotate along the torus so that we have a rotating ring lattice, with the velocity along the ring specified by $v$. The way that we proceeded in Ref.~\cite{WAN07} is probably  uncontroversial, but a closer examination reveals a number of subtleties.

To begin with, the notion of an explicitly time dependent potential energy is not part of the standard edifice of classical mechanics, and so not part of quantum mechanics either. We resolve this immediate issue with the Galilean invariance principle that in a frame moving with the potential energy at the uniform speed $v$, the classical one-particle physics is the same as in the stationary frame when the potential does not move. In terms of the usual classical position and conjugate momentum variables, the Hamiltonian is
\beq
H_v = \frac{p^2}{2m} + V(x)\,.
\eeq
We may quantize the Hamiltonian in the usual way by replacing the position and momentum variables with the quantum operators $\hat x$ and $\hat p$, to have
\beq
\hat H_v =  \frac{\hat p^2}{2m} + V(\hat x)\,.
\eeq

Interestingly,  the canonical commutator does not uniquely determine the quantized momentum operator $\hat p$. For instance, if $[\hat x,\hat p] = i\hbar$, then $[\hat x,\hat p + f(\hat x,t)] = i\hbar$ also holds true for an arbitrary function $f(x,t)$. For our purposes it suffices to set $f(x,t)=-\lambda$, a so far undetermined constant. In position representation, in the frame moving with the lattice, we therefore write the quantum Hamiltonian as
\beq
H_v = \frac{1}{2m} \left(\frac{\hbar}{i}\,\frac{\partial}{\partial x}-\lambda\right)^2 + V(x)+K\,.
\eeq
Here we have added another constant $K$ that has no effect on the dynamics, but is at our disposal for later convenience.

Let us now transform from the moving coordinate system to the stationary frame. The coordinate transformation, a Galilean transformation, is defined by 
\beq
\tau = t,\quad x = \xi + v \tau\,,
\label{GALTRF}
\eeq
where $\xi$ and $\tau$ are position and time in the stationary lab frame. It follows from the time dependent Schr\"odinger equation in the rotating frame for the wave function $\psi_v(x,t)$ that the corresponding stationary-frame wave function
\beq
\psi(\xi,\tau) = \psi_v(\xi+v\tau,\tau) 
\eeq
also satisfies the time dependent Schr\"odinger equation with the Hamiltonian
\bea
H &=& \frac{1}{2m}\left[\frac{\hbar}{i}\,\frac{\partial}{\partial \xi}-(\lambda-mv)\right]^2 + V(\xi-v\tau)\nonumber\\
&&+ v\lambda-\half\,mv^2 + K\,.
\eea

The crux of the present argument is as follows: In the stationary frame, and in the absence of the moving potential ($V\rightarrow0$), we want to regain the ordinary Schr\"odinger equation for a free particle, with the usual free-particle Hamiltonian
\beq
H = -\frac{\hbar^2}{2m}\,\frac{\partial^2}{\partial \xi^2}
\label{FREEHAM}
\eeq
and periodic boundary condition for the wave function $\psi(\xi,\tau)$ in the variable $\xi$.
This happens only if we choose
\beq
\lambda = mv,\quad K = -\half mv^2\,.
\eeq
This puts the rotating-frame Hamiltonian into the form
\beq
H_v = \frac{1}{2m} \left(\frac{\hbar}{i}\,\frac{\partial}{\partial x}-mv\right)^2 + V(x) - \half\,mv^2\,.
\label{HV}
\eeq
The rotating-frame wave function $\psi_v(x,t)$ has periodic boundary conditions in the variable $x$.

The Hamiltonian~\eq{HV} is still nonstandard form. To remedy this situation, we introduce a transformation of the wave function $\psi_v\rightarrow\Psi$, a momentum translation, by
\beq
\psi_v(x,t) = e^{i\frac{mvx}{\hbar}}\,\Psi(x,t)\,.
\label{psiToPsi}
\eeq
Upon this unitary transformation the Hamiltonian becomes
\beq
H_v' = -\frac{\hbar^2}{2m}\,\frac{\partial^2}{\partial x^2} + V(x) - \half\,mv^2\,,
\label{HVP}
\eeq
and the wave function $\Psi$ acquires {\em twisted\/} boundary conditions,
\beq
\Psi(x+L,t) = e^{-i\frac{mvL}{\hbar}}\,\Psi(x,t)\,.
\eeq

Our development has introduced three frames: The stationary lab frame (LF), the rotating frame (RF), and the momentum translated rotating frame (MTRF). In quantum mechanics the frames are related by unitary transformations. By choice, the bulk of our analysis is in the MTRF, but all three frames figure in the argument and have to be distinguished carefully.

We illustrate the distinction between these frames by solving for the stationary states in the absence of any external potential, $V=0$, in the MTRF. The process is standard. In particular, the operator $\frac{\hbar}{i}\,\frac{\partial}{\partial x}$ is hermitian with respect to the inner product~\eq{INNERPROD} even with the MTRF twisted boundary conditions. The normalized stationary states, solutions to the time dependent Schr\"odinger equation for which all quantum mechanical expectation values are independent of time, read
\beq
\Psi_n(x,t) = \frac{1}{\sqrt{L}}\,e^{i (K_n x - \Omega_n t)}\,,
\label{MTRFWF}
\eeq
with
\bea
K_n &=& k_n - k_v;\nonumber\\k_v &=& mv/\hbar;\quad k_n = \frac{2\pi n}{L},\,n=0,\pm1,\ldots;\nonumber\\
\Omega_n &=& \frac{\hbar}{2m}(K_n^2 - k_v^2)\,.
\label{ROTSOL}
\eea
To go  from the MTRF to the LF we apply both the appropriate momentum translation and the inverse of the Galilean transformation~\eq{GALTRF}, and find the stationary states
\beq
\psi_n(\xi,\tau) = \frac{1}{\sqrt{L}}\, e^{i(k_n \xi - \omega_n \tau)},\quad \omega_n = \frac{\hbar k_n^2}{2m}\,.
\label{STATSOL}
\eeq
These are the usual stationary states of the LF Hamiltonian~\eq{FREEHAM}, and satisfy the proper LF periodic boundary conditions.

Simple as this exercise is, it already underscores the problem that one encounters in connection with statistical mechanics. Namely, the energies of the states in the MTRF and in the LF are different, $\hbar\Omega_n$ and $\hbar\omega_n$, and the difference is not just a constant. What would be the thermal equilibrium, and even the ground state that gets prepared at zero temperature, is therefore ambiguous. The question becomes even more acute if there is a potential present, because  the Hamiltonian then depends explicitly on time in the stationary frame and standard statistical mechanics  is inapplicable. We will not attempt to clean up these issues, but simply hypothesize a resolution as we go along.

We next proceed to Hubbard type models for a lattice. Here and in the rest of the paper we restrict ourselves to the basic one-band (tight-binding) model, and ignore altogether the possible coupling of the energy bands~ that may occur with increasing strength of atom-atom interactions~\cite{DUA05,BUC10}.

We begin in the MTRF, Hamiltonian~\eq{HVP}, with what is essentially a rederivation of Bloch's theorem. Thus, let us introduce the lattice translation operator $U$ and its inverse $U^{-1}$ via their actions on the wave function by
\beq
(U\Psi)(x) = \Psi(x+a),\quad (U^{-1}\Psi)(x) = \Psi(x-a)\,.
\eeq
This operator preserves the inner products, $(U\Psi,U\Phi)=(\Psi,\Phi)$, and is invertible, so it is unitary. Moreover, for a periodic potential $V(x)$ it commutes with the Hamiltonian $H'_v$ of Eq.~\eq{HVP}. Therefore $H'_v$ and $U$ can be diagonalized simultaneously. Let $\Psi$ be a simultaneous eigenfunction and $\lambda$ the corresponding eigenvalue of $U$. By virtue of the twisted boundary conditions we have
\beq
(U^N \Psi)(x) = \lambda^N \Psi(x) = \Psi(x+L) = e^{-ik_vL}\Psi(x)\,.
\eeq
The possible eigenvalues of $U$ are therefore of the form
\beq
\lambda_n = e^{i K_na}\,,
\eeq
see Eqs.~\eq{ROTSOL} for the notations used in the present argument.
By writing an eigenvalue and eigenvector corresponding to $K_n$ in the form
\beq
\Psi_n(x) = e^{iK_n x} u_n(x)\,,
\label{TWISTBLOCH}
\eeq
we immediately see that $u_n(x)$ has to be periodic over the distance $a$. This is Bloch's theorem for twisted boundary conditions.

Obviously, by virtue of the unitary transformation~\eq{psiToPsi},  we would write Bloch's theorem in the RF, for the Hamiltonian~\eq{HV}, in the form
\beq
\psi_n(x) = e^{ik_n x} u_n(x)\,,
\eeq
with the same functions $u_n(x)$ as in Eq.~\eq{TWISTBLOCH}.

When the lattice potential is weak, the rotation may rearrange the band structure. However, as we are interested in the tight-binding limit, we assume that the lowest energy band remains essentially undeformed by rotation, and consider only the $N$ states therein.

Let us next define the analogs of Wannier functions, given the twisted boundary conditions:
\bea
W_n(x) &=& \frac{1}{\sqrt N} \sum_{n'} e^{-i k_{n'}x_n} \Psi_{n'}(x)\nonumber\\
&=&  \frac{1}{\sqrt N} \,e^{-i k_v x} \sum_{n'} e^{i k_{n'}(x-x_n) }u_{n'}(x)\nonumber\\
&=& \frac{1}{\sqrt N} \,e^{-i k_v x} \sum_{n'} e^{i k_{n'}(x-x_n) }u_{n'}(x-x_n)\nonumber\\
&=& e^{-i k_v x}  w(x-x_n)\\
&=& e^{-i k_v x_n} W(x-x_n)\label{WTR}\,,
\eea
where
\beq
w(x) =\frac{1}{\sqrt N} \sum_n e^{ik_n x} u_n(x)
\label{NONTWISTWANNIER}
\eeq
is evidently the Wannier function in the RF for the lattice position $n=0$, and 
\beq
W(x) = e^{-i k_v x} w(x)
\eeq
likewise in the MTRF. Incidentally, an attempt to construct Wannier functions numerically quickly reveals that they are not unique. Namely, the Bloch functions~\eq{TWISTBLOCH} can be multiplied by arbitrary phase factors to give equally good Bloch functions, but possibly completely different Wannier functions. Our usual choice is to put a minimum of the lattice potential at $x=0$ and pick the Bloch states so that $\Psi_n(0)$ is real and positive. Empirically, this choice seems to produce Wannier functions $W_n(x)$, each of which is narrowly centered at $x_n$.  Whatever the choice, according to Eq.~\eq{WTR} the Wannier functions $W_n(x)$ are translated copies of a single function $W(x)$. The functions $W_n(x)$ are orthonormal if the original Bloch states~\eq{TWISTBLOCH} are, and in fact make an orthonormal basis for the states in the lowest energy band.

The MTRF Wannier functions $W_n(x)$ and $W(x)\equiv W_0(x)$ themselves obey twisted boundary conditions by construction, and so do all of their linear combinations. However, Eq.~\eq{WTR} also suggests another mechanism whereby twisted boundary conditions may be satisfied. Namely, a linear combination of functions of this form with an infinite number of coefficients $c_n$ satisfying periodic boundary conditions, $c_n = c_{n+N}$, if it exists in the first place, satisfies twisted boundary conditions {no matter what the function\/} $W(x)$ is. Conversely, if some $W(x)$ were strictly localized around $x=0$, equal to zero by $|x|\ge L/2$, and did not repeat at the intervals of $L$, then the only way to make a wave packet satisfying {\em twisted\/} boundary conditions from the functions $W_n(x) = e^{-ik_vx_n}W(x-x_n)$ would be to insist on {\em periodic\/} boundary conditions for the expansion coefficients $c_n$.

In short, the following modeling suggests itself:
\begin{itemize}
\item[(i)] We adopt a function $W(x)$ that is localized around $x=0$, without periodic recurrences, and in such a way that $W_n(x) = e^{-ik_vx_n} W(x-x_n)$ for different $n$ may be reasonably  regarded as an orthonormal set of functions.
\item[(ii)] We consider only linear combinations of the functions $W_n(x)$ of the form
\bea
\Psi(x) &=& \sum_{n=-\infty}^{\infty} c_n W_n(x)\nonumber\\
&=& \sum_{n=-\infty}^{\infty} c_n \,e^{-ik_vx_n} W(x-x_n)\,,
\label{WFC}
\eea
where the expansion coefficients satisfy periodic boundary conditions, $c_n = c_{n+N}$.
\item[(iii)]  The wave function $W_n(x)=e^{-ik_vx_n}W(x-x_n)$ is the one-particle state representing an atom that resides at the site $n$, and $b_n$ is the corresponding boson operator.
\end{itemize}

The function $W(x)$ specifying the Wannier functions restricted to the interval $[-L/2,L/2)$ provides an example of a function of this kind. Specifically, given the Wannier function around $x=0$, $W_0(x)$, and the unit step function $\theta(x)$, we could define $W(x) = W_0(x) \theta(x+L/2)\theta(L/2-x)$. The ensuing construction of the lattice states is then precisely the same as if we simply considered the original Wannier function restricted to a proper finite set of indices such as $n\in[-N/2,N/2-1)$. Likewise, one could use an approximation to the ground state of a particle in the potential well at $x=0$ to model the function $W(x)$. Such constructions are most useful if the dependence on rotation is primarily contained in the prefactors $e^{-ik_vx_n}$ of the basis wave functions $W_n(x)$. After we have adopted these one-particle states and boson operators, the lattice per se is periodic, with the sites $n$ and $n+N$ regarded as the same.

Given the functions $W_n(x)$, we finally set up the corresponding Hubbard model with nearest-neighbor tunneling and atom-atom interactions in the usual manner,
\beq
\frac{H}{\hbar}\! =\!\frac{1}{2}\sum_n\left[\! -(e^{i\phi}  Jb^\dagger_{n+1} b_n \! +\! e^{-i\phi}J^*b^\dagger_nb_{n+1}) \!+ \!U b^\dagger_n b^\dagger_n b_n b_n\right]\!\!.
\label{POSHAM}
\eeq
Logic would dictate that we use the notation $H'_v$ for this MTRF operator, but we write $H$ anyway.
We have
\bea
J &=&- \frac{2}{\hbar}\int_{-L/2}^{L/2} dx\,W^*(x-a)(H'_vW)(x)\\
&=&- \frac{2}{\hbar}\,e^{-i\phi}\int_{-L/2}^{L/2} dx\,w^*(x-a)(H_vw)(x)\,.
\label{JFORM}
\eea
Since the periodic boundary conditions of the MTRF are awkward to deal with, using the transformation~\eq{psiToPsi} we have also expressed the tunneling matrix element  in terms of what would approximate the Wannier functions for the RF Hamiltonian~\eq{HV}.  The phase factor $\phi$ equals the phase twist per lattice site, or
\beq
\phi = \frac{\Phi}{N},\quad \Phi = \frac{mvL}{\hbar}\,,
\eeq
where $\Phi$ is the end-to-end phase twist over the lattice.
The atom-atom interaction part depends on the $s$-wave scattering length $a_0$, but to obtain it accurately we need to know also about the structure of the wave function in the directions transverse to $x$. Given a full 3D form of the Wannier functions, we could write
\beq
U = \frac{4\pi \hbar a_0}{m}\int d^3x\,|W({\bf x})|^4\,.
\eeq
A periodic lattice is implied, so that the sites $n=0$ and $n=N$ are the same. We have finally dropped the constant $-\half mv^2$ in the Hamiltonian as it has no effect on either dynamics or thermodynamics.

We also need the Hamiltonian in lattice momentum representation. Specifically, let us define the boson operators in lattice momentum space $B_q$ in such a way that
\beq
b_n = \frac{1}{\sqrt N}\sum_q e^{iqn} B_q\, \Leftrightarrow\, B_q = \frac{1}{\sqrt N} \sum_n e^{-iqn} b_n\,,
\label{MOMHAM}
\eeq
then the Hamiltonian reads
\bea
H &=& -J\sum_q \cos(q-\phi)\,B^\dagger_qB_q\nonumber\\
&&+\frac{U}{2N} \sum_{q_1,q_2,q_3,q_4}\delta_{q_1+q_2,q_3+q_4} B^\dagger_{q_1}B^\dagger_{q_2}B_{q_3}B_{q_4}\,.
\label{MRH}
\eea
With definitions in terms of the lattice spacing $a$ such as $q\equiv ak_n$, the discrete numbering of the lattice momenta is hidden and the lattice momenta $q$ are made dimensionless. Moreover, by virtue of the periodicity in lattice momentum space, sums and comparisons of lattice momenta are modulo $2\pi$ unless there is an explicit reason to proceed differently. If the original operators $b_n$ annihilate Wannier states of the lattice, the operators $B_q$ annihilate Bloch states.

Several items remain to be cleaned up. First, some fine tuning on the rotation phase is in order. Suppose that we have at the origin the harmonic oscillator potential $V(x) = \half m\omega^2 x^2$, and use the ground state following from the RF Hamiltonian $H_v$, Eq.~\eq{HV} as the ``Wannier function'' $w(x)$ in Eq.~\eq{JFORM}.  The tunneling matrix element is then
\beq
J = \omega\,e^{-\frac{m\omega a^2}{\hbar}}\,,
\eeq
independent of the rotation phase $\phi$. On the other hand, when we use a sinusoidal potential $V(x)$ as appropriate for an ideal optical lattice in a numerical solution of the RF Hamiltonian~\eq{HV}, we find that both the magnitude and the phase of the tunneling amplitude $J$ depend somewhat on $\phi$. In particular, there is a phase drag of sorts incorporated into $J$. However, in the tight-binding limit these are small corrections. Given that our model is heavy-handed in many other details anyway, we will assume that the phase $\phi$ accounts for all effects of the rotation of the lattice.

Second, one could augment each of the functions $W_n(x)$ with an arbitrary phase factor $e^{-i\varphi_n}$,
\beq
W_n(x)\rightarrow e^{-i\varphi_n}W_n(x)\,.
\label{PHADJ}
\eeq
 Indeed, this operation does not change the orthonormality of the functions $W_n(x)$, or the function space they span. The end result would be additional phase factors in the site-to-site tunneling amplitudes, such as
\beq
e^{i\phi}  Jb^\dagger_{n+1} b_n \rightarrow e^{i(\phi+\varphi_{n+1}-\varphi_{n})}  Jb^\dagger_{n+1} b_n\,.
\label{ROTPHADJ}
\eeq
The lattice remains periodic, wraps around at $n=N$, only if the added phase factors wrap around, so that we have $e^{i\varphi_{n+N}} = e^{i\varphi_{n}}$. This means that the phases added to the tunneling matrix elements,  $\varphi_{n+1}-\varphi_n$, must add up to an integer multiple of $2\pi$. Since the absolute phases assigned to the members of an orthonormal basis cannot have any effect on the physics, the total transformation~\eq{PHADJ} and~\eq{ROTPHADJ} to alter the phase winding cannot change the physics of the lattice in any way.

Because the effect of the transformation of Eqs.~\eq{PHADJ} and~\eq{ROTPHADJ} on the Hamiltonian~\eq{POSHAM} with the choice $\varphi_n = n\Delta\phi/N$ is exactly the same as the change of the rotation phase by $\Delta\phi$, one might be tempted to conversely surmise that the physics repeats periodically when the rotation speed is increased; is unchanged whenever $\Phi$ changes by $2\pi$, or the rotation phase $\phi$ changes by $2\pi/N$. This notion, if valid, would severely limit the possibilities to control the lattice by varying the rotation speed. Fortunately, it is incorrect. 

The transformation~\eq{PHADJ} comes with both phase changes in the basis functions and an apparent change in the rotation phase~\eq{ROTPHADJ}  that cancel inasmuch as physical observables are concerned. On the other hand, a change in the rotation speed and the accompanying changes in the Hamiltonian do not per se alter the basis functions. The spectrum of any observable, such as the Hamiltonian, is periodic in $\Phi$ with the period of $2\pi$ since the absolute phases of the basis wave functions used in the calculations have no effect on the diagonalization. However, if we were to change the rotation phase adiabatically so that $\Phi$ changes by $2\pi$, the wave function of an energy eigenstate could change, and along with it measurable properties of the system. The situation is analogous to what happens to a function with a branch cut starting from the origin of the complex plane: moving the complex argument of the function a full circle about the origin could put the value of the function to another branch. 

For instance, according to~\eq{PHADJ} and~\eq{ROTPHADJ} one could effectively flip the sign of $J$ in the Hamiltonian by flipping the sign of every other basis function without any physical consequences, but only if the number of the lattice sites is even. Otherwise the phase adjustments in the Hamiltonian over the lattice would not add up to an integer multiple of $2\pi$. This tells us that the physics may depend on the parity of the number of lattice sites. On the other hand, one could effectively flip the sign of $J$ and adjust the energy spectrum accordingly by spinning the lattice up and thereby by adding $\pi$ to the rotation phase, but it is not obvious without further investigation what happens to the observable properties of the lattice.

\section{One-atom states}\label{ONEATOM}
We begin our investigation of the physics due to the rotation with the case of one atom in the lattice. If it were possible to extinguish atom-atom interactions, for instance with the aid of a Feshbach resonance, many bosons may be put in the same one-particle state without any side effects. Such a many-atom sample could aid in the observation of the phenomena we will discuss.

In the case of one particle the MTRF Hamiltonian~\eq{MRH} is diagonalized trivially. Each lattice momentum eigenstate annihilated by the boson operator $B_q$ is also an eigenstate of energy with the characteristic frequency
\beq
\omega_q = -J \cos(q-\phi)\,.
\eeq
The corresponding energy eigenstates are given by $\ket{\Psi_q} = B^\dagger_q\ket0$, or, in the first-quantized representation in terms of the Wannier-like function $W$,
\beq
\Psi_q({x}) \equiv \Psi(q,\phi;x) =  \frac{1}{\sqrt{N}}\sum_n e^{i(q-\phi)n}W(x-x_n)\,.
\label{NONINTST}
\eeq
\commentout{
The second form emphasizes the interlinked dependence on lattice momentum $q$ and rotation phase $\phi$. Let us assume that the value of $\phi$ may be arbitrary. Nonetheless, it is uniquely decomposed into the form
\beq
\phi = q_\phi + \bar\phi\,,
\eeq
where $\bar\phi$ resides in the interval $[-\pi/N,\pi/N]$ and $q_\phi$ is in form a legal lattice momentum. Then we also have
\beq
\Psi(q,\phi;x) = \Psi(q-q_\phi,\bar\phi;x)\,,
\label{QPHIMIX}
\eeq
and in the latter form the rotation phase is restricted to the interval $[-\pi/N,\pi/N]$. In this way one may trade rotation phase for lattice momentum, or the other way round. The lattice momenta we have quoted in these arguments  are in a frame that includes the momentum translation~\eq{psiToPsi}.

Lattice momenta can be measured experimentally~\cite{OTT04,WIN06}. For an ordinary stationary linear lattice this is done by turning the lattice off in a time that is long enough that structure of the states below the length scale of lattice spacing has time to smooth out, but short enough that the structure of the wave function survives on the length scale of the lattice. The effect is to convert lattice momentum residing in the first Brillouin zone to ordinary momentum. We will now have to figure out how the corresponding process would work out in the ring lattice.

There is another deep subtlety here. Namely, unitary transformations do not change quantum mechanics per se, but you need to make a corresponding change of the states and operators. For instance, if momentum operator is $\hat p = \frac{\hbar}{i}\,\frac{\partial}{\partial x}$ in some unitary representation of quantum mechanics, then it may look different in another unitary representation. Conversely, if we were to insist that $\hat p = \frac{\hbar}{i}\,\frac{\partial}{\partial x}$ is the momentum operator, we will somehow have to find out in which unitary representation of quantum mechanics this this statement is true. In short, we will have to decide on a preferred unitary representation of quantum mechanics. Given that we made the transition from classical mechanics to quantum mechanics by employing position  representation a frame rotating with the lattice, we use the position representation in the rotating frame as our preferred representation of quantum mechanics.}

Let us now consider measurement of lattice momentum in a ring lattice by a variation of the methods that were successfully applied to measure lattice momenta in usual linear lattices~\cite{OTT04,WIN06}. Imagine first that the ring lattice were straightened out, somehow in such a way that the boundary conditions etc.\ still work out as in the closed ring. The wave function of the state $\Psi_q(x)$ is given by Eq.~\eq{NONINTST} in the MTRF, so we undo the momentum translation and in the RF find
\bea
\psi_q(x) =
\frac{1}{\sqrt{N}}\sum_n  e^{i q n}[e^{i\phi(x-x_n)/a}W(x-x_n)]\,.
\eea
The wave function inside the square brackets is localized to a fraction of the lattice spacing $a$, and as such should have little effect on lattice momenta over the scale of the first Brillouin zone. Specifically, assume that the lattice is turned off so slowly that the details of the wave function on the scale of lattice spacing $a$ get smoothed out, but fast enough that the structure  over the length scale of the entire lattice $L$ survives~\cite{OTT04,WIN06}. With $n\equiv x/a$, the wave function then becomes
\beq
\psi_q \propto e^{iqx/a}\,.
\eeq
Now, given the RF momentum operator as shown in the Hamiltonian~\eq{HV}, we have
\beq
\left(\frac{\hbar}{i}\,\frac{\partial}{\partial x}-mv\right)\psi_q = \left(\frac{\hbar q}{a}-mv\right)\psi_q\,,
\eeq
which identifies the measured kinetic momentum, mass times velocity, in the moving frame. Finally, applying the Galilean transformation, we see that in the LF the momentum measured in this way would be $\hbar q/a$, simply the lattice momentum $q$ translated into units of momentum. Therefore, as in the case with an ordinary stationary linear lattice, the lattice momentum $q$ may be converted to linear momentum and measured.

In practice an operation that just straightens out a ring lattice appears implausible. Instead, think of a section of the ring that looks approximately like a straight piece. The atom, when released from different sections would have different directions of momenta, but still the same speed and energy. The latter could then be measured. However, as design of experiments is not our aim, we will not pursue a more quantitative analysis.

\commentout{
Finally, at a given fixed time the lab frame wave function is just a translated version of the rotating-frame function. But the measurements of lattice momentum are typically based on interference of the waves released from given lattice sites, and the interference works exactly the same whether the lattice 
}
We have considered three different frames in our derivation: stationary lab frame LF, rotating frame RF, and momentum translated rotated frame MTRF. It may come as a surprise that, after we undo the momentum translation and are still formally in the RF, the lattice momentum $q$ is the lattice momentum that would be observed in the LF. Nevertheless, an inspection of the Hamiltonian~\eq{HV} shows why this is  consistent: To get to the RF, we subtract the momentum due to the motion of the frame, $mv$, from the quantum mechanical momentum operator $\hat p = \frac{\hbar}{i}\frac{\partial}{\partial x}$, which means that $\hat p$ should represent the momentum in the stationary LF. All of this is consistent with the observations surrounding Eqs.~\eq{MTRFWF} --\eq{STATSOL}.

On the other hand, when we quote values of energies, they are always in a moving frame, RF or MTRF. Momentum translation has no effect  on energy, as it comes with canceling transformations of both the state and the energy operator (Hamiltonian).

\begin{figure}
\includegraphics[width=8.0cm]{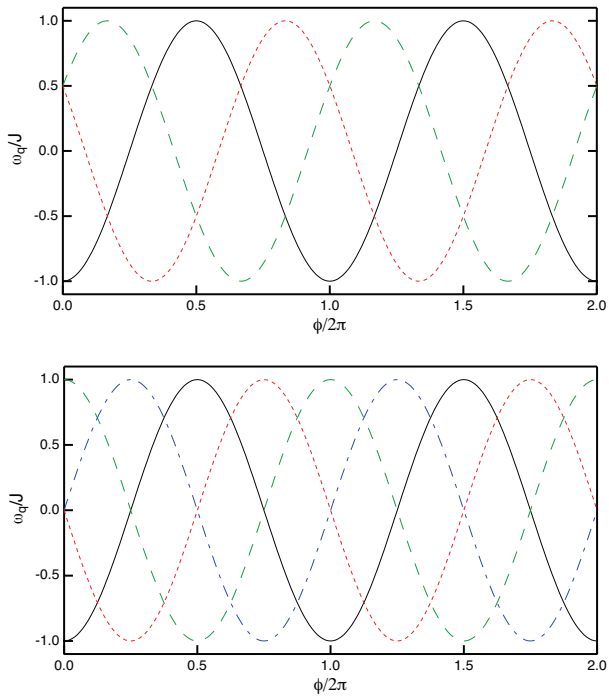}
\caption{Energies of the lattice eigenstates for $N=3$ (upper panel) and $N=4$ (lower panel) sites as a function of the site-to-site rotation phase $\phi$. The graphs are for lattice momenta $q=0$ (solid black line), $q=2\pi/N$ (red dotted line), $q=4\pi/N$ (green dashed line) and $q=6\pi/N$ (blue dash-dotted line).}
\label{ENFIG}
\end{figure}

We plot in Fig.~\ref{ENFIG} the characteristic frequencies of all energy eigenstates as a function of the rotation phase $\phi$ for a lattice with $N = $ 3 and 4 sites, upper and lower panels. There is a qualitative difference in the spectra in accordance with the observation that the parity of the number of states may make a difference. Since the plot presents $\omega_q/J$, for $J>0$ the energy increases from bottom to top on the vertical axes.

Suppose first that, in fact, $J<0$ holds true, so that lowest energies are at the top. Without rotation the ground state of a lattice with an odd number of sites is doubly degenerate. Moreover, the degenerate states have $q\ne 0$, i.e., present a flow along the lattice. If we prepare the ground state and measure the lattice momentum, we find one of two nonzero values at random. If the lattice has an even number of sites, the ground state is nondegenerate. It has the lattice momentum $\pi$, which is physically equivalent to $-\pi$. Depending on the measurement scheme the result may be $\pi$ or $-\pi$, but they count as the same.

On the other hand, for $J>0$ the ground state is nondegenerate and has zero lattice momentum, no matter if the number of lattice sites is even or odd. The sign of the hopping matrix elements may have dramatic qualitative consequences if the number of lattice sites is odd.

Incidentally, in the examples we have studied analytically and numerically, we always found $J>0$. There may be a general principle dictating that for the kind of  passive tunneling we have considered $J>0$ should  hold true. But if so, active schemes that control the tunneling externally, say, using light-induced Raman transitions, are clearly not restricted to $J>0$.

As we have already noted, the question of the ground state is precarious once the lattice rotates. Let us posit that preparation to a very low temperature produces the lowest-energy state in the rotating frame. We also assume  $J>0$ for the sake of the argument, and take the example with $N=4$. Consultation of the lower panel in Fig.~\ref{ENFIG} then shows that, by varying the phase $\phi$ over a $2\pi$ interval, for each $q$ a region of $\phi$ emerges such that the corresponding state $q$ is the ground state and gets prepared. For instance, $q=0$ is a ground state in the interval $\phi/2\pi\in[-1/8,1/8]$, $q=\pi/2$ in the interval $\phi/2\pi\in[1/8,3/8]$, and so on. 
Such switching of the nature of the ground state in a rotating ring lattice as a function of the rotation speed has been noted before, and in the many-atom case with weak atom-atom interactions it may lead to a Schr\"odinger cat superposition of macroscopic flow states~\cite{HAL06,REY07}. 

Thermal preparation presents an example about the roles of the rotation phases and system observables. Consider the ground state first around $\phi=0$, which in the MTRF reads
\beq
\Psi_0(x) = \frac{1}{\sqrt{N}}\sum_ne^{-i\phi n}W(x-x_n)\,.
\label{NRGRD}
\eeq
On the other hand, the ground state around $\phi=\pi/2$ has $q = \pi/2$, and the wave function can be written, among others, in the form
\beq
\Psi_{\pi/2}(x) =  \frac{1}{\sqrt{N}}\sum_ne^{-i(\phi -\pi/2)n}W(x-x_n)\,.
\eeq
This is the same function of $x$ as~\eq{NRGRD}, albeit with the replacement $\phi\rightarrow\phi-\pi/2$. As a function of $\phi$ and $x$, the ground state for all $\phi$ is the same as the ground state in the interval $\phi/2\pi\in[-1/8,1/8)$ repeated in $\phi$ with the period $2\pi/N=\pi/2$. Nevertheless, even though in the MTRF the wave function repeats with the period $\Delta\phi=2\pi/N$, this is not necessarily the case with the observed quantities. In fact, lattice momentum as measured in the LF steps by $2\pi/N$ every time the $q$ label of the ground state switches. This would probably be the educated guess of most colleagues familiar with vortices in trapped gases.

Things are different in an interesting way if one considers ``slow'' time dependent variation of the rotation velocity. We call such variation adiabatic. At this stage there are no interactions between the energy eigenstates and they (Bloch states) belong to different discrete symmetries, namely, eigenvalues of the lattice translation operator $U$. We therefore assume that they are extremely robust even if a degeneracy is crossed when the rotation speed is varied in time. Put differently, when two states cross, they have a difference of an integer multiple of $2\pi$ in their end-to-end phase winding. A continuous dynamics cannot discontinuously change the phase winding, so that transitions between the states are not possible.  In Fig.~\ref{ENFIG} one would follow a curve of a given color with varying $\phi$, and the lattice momentum observed in the LF is independent of the rotation speed of the lattice.

It might seem peculiar that in a lattice an energy eigenstate such as the ground state  cannot be wound up adiabatically, but this appears to be a manifestation of the same phase rigidity that sustains a persistent current in a ring. We argued a while ago~\cite{JAV98} that one has to cut the superfluid inside an essentially one-dimensional ring and thereby severe the continuity of the phase if one is to alter the state of circulation. The lattice potential cuts the ring, but not all the way. If the rate of change of the rotation phase $\Phi$ is sufficiently small compared to the tunneling matrix element $J$, the ``fluid'' of the single atom can still adjust and respond as if it were continuous.

Continuing from the preceding ground-state example, one could first prepare the ground state $\Psi_{\pi/2}(x)$ in a rotating lattice corresponding to the rotation phase $\phi=\pi/4$ and then wind down the rotation adiabatically while the state $\Psi_{\pi/2}(x)$ stays this way. This is a way to prepare in principle any eigenstate of the stationary lattice.

As one more item of the phenomenology of the rotating lattice, let us discuss the fate of a wave packet. We use group velocity as the tool. The concept of group velocity is increasingly useful, the larger the number of lattice sites and the broader  the wave packet (so that the role of dispersion is diminished). We will not carry out the associated quantitative analysis for finite-size lattices or wave packets, but simply assume that group velocity yields useful qualitative predictions. It is
\beq
v_g = \frac{d\omega_k}{dk} = a J \sin(q-\phi)\,,
\label{GRVEL}
\eeq
where, in order to permit a direct comparison with the rotation velocity, we have included the lattice constant $a$ explicitly, as in $k=q/a$. Group velocity governs the evolution of the spatial envelope of the wave packet, and is given by the expression~\eq{GRVEL} in both the RF and the MTRF.

Take a localized stationary wave packet with $q\simeq0$ prepared when the lattice does not rotate, and assume that the lattice is then set in motion adiabatically in such a way that the decomposition of the wave packet into its component $q$ states is not  perturbed by the process. In the rotating frames the wave packet therefore picks up the group velocity
\beq
v_g = - a J \sin\phi \simeq - \frac{1}{N}\,\frac{ ma^2 J}{\hbar}\, v\,,
\eeq
where the latter form applies for a small rotation velocity. The group velocity is small compared to the rotation velocity, both because of the factor $1/N$ and because of the second factor that is essentially the ratio of the tunneling frequency to the photon recoil frequency associated with an atom in the optical lattice. This means that the wave packet starts moving approximately, but not exactly, with the rotating lattice. Depending on the sign of the transition matrix element $J$, the wave packet either slightly lags or may even lead the lattice. We emphasize the curious contrast: lattice momenta are unchanged when the lattice is spun up, but the wave packet basically tracks the moving lattice.

We have described several peculiar phenomena that should occur with a single particle or noninteracting particles in a rotating ring lattice. Some of them, such as the scheme to prepare an arbitrary eigenstate of energy, were based on nontrivial assumptions. But then, we can turn the tables and say that an experiment would test the validity of these assumptions.

\section{Lattice dimer}\label{TWOATOMS}
We next proceed to the case of two interacting atoms in the lattice. We have discussed an analogous situation in detail before~\cite{JAV10}, but some reorientation is in order here. While our emphasis was on the limit of an infinitely long lattice and the periodic boundary conditions were a matter of convenience,  in a laboratory ring lattice the topology of the ring is, and a small number of sites could be, a physical reality, and may necessitate a numerical solution of the system. For the most part, however, our main emphasis is on the rotation phases, and when possible we piggyback on our earlier analysis~\cite{JAV10} of lattice dimers.

Thus, we write the most general state of two bosons in the lattice in the lattice momentum representation as
\beq
\ket\psi = \sum_q A(q) B^\dagger_{\half P+q}B^\dagger_{\half P-q}\ket 0,
\label{ASTATE}
\eeq 
where $\ket0$ is  the vacuum with no atoms present. Here $P$ is the total lattice momentum of the dimer of sorts, the value of the conserved quantity
\beq
\hat P = \sum_q q\, B^\dagger_q B_q\,.
\eeq
We let the value $P$ of the operator $\hat P$ range from $-2\pi$ to $2\pi$, so that $\half P$ runs over the usual interval $[-\pi,\pi)$ of lattice momenta. Also, $\half P$ need not be a valid lattice momentum, but $\half P \pm q$ have to be. This means that in the sum the lattice-momentum like quantity $q$ may either run over legal ``integer'' lattice momenta, or it may be displaced from legal lattice momenta by a half-step $\pi/N$. This is what we earlier termed ``half-integer'' lattice momenta. Either way, the sum over $q$ in Eq.~\eq{ASTATE} runs over a set of $N$ values so that $\half P + q$ and $\half P - q$ both run once over all permissible lattice momenta, with no two values separated by $2\pi$ or more.

The coefficients $A(q)$ govern the internal structure of the dimer. They remain to be determined. By the boson symmetry they can be, and are, chosen so that $A(q)=A(-q)$. For convenience we regard the coefficients $A(q)$  as periodic over $2\pi$. The states associated with the coefficients $A(q)$ and $A(-q)$, $B^\dagger_{\half P\pm q}B^\dagger_{\half P\mp q}\ket 0$ are the same. As a result, the usual inner product of the states of the lattice system is expressed in terms of the expansion coefficients as
\beq
(\psi_A,\psi_B) = 2 \sum_q A^*(q) B(q)\,.
\eeq
An explicit solution to the problem of a few-site lattice has to be tailored to the even/odd number of lattice sites and integer/half-integer lattice momenta, with different sets of  possible values of the lattice momenta $q$~\cite{JAV10}, but we will not embark on an enumeration of the various cases.

We now turn to the specifics of a rotating ring lattice. In the MTRF the time independent Schr\"odinger equation gives an equation for the coefficients $A(q)$
\beq
-2J \cos(\half P - \phi) \cos q\,A(q) + \frac{U}{L}\sum_{q'} A(q') = \frac{E}{\hbar} A(q)\,.
\eeq
Defining the overall frequency scale $\Omega(P,\phi)$ as
\beq
\Omega(P,\phi) = 2 J \cos(\half P - \phi)\,,
\label{OMEGADEF}
\eeq
and dimensionless variables representing energy and the strength of atom-atom interactions
\beq
\omega = \frac{E}{\hbar\Omega},\quad {\cal K} = \frac{U}{\Omega}\,,
\eeq
the energy eigenvalue problem may be written
\beq
\frac{1}{N} \sum_q \frac{1}{\omega + \cos q} = \frac{1}{\cal K}\,.
\eeq
The corresponding unit-normalized energy eigenstates are defined by the coefficients
\beq
A(\omega,q) = \frac{C(\omega)}{\omega + \cos q},\, C(\omega) = \left[ \sum_q \frac{2}{(\omega+\cos q)^2}\right]^{-1/2}\,.
\eeq

We~\cite{JAV10,SAN11,ODO11} and others~\cite{ORS05,WIN06,GRU07,VAL08} have discussed the nature of the solutions before. There is a band of  continuum states restricted to the interval of characteristic frequencies $(-|\Omega|,|\Omega|)$,  and one  bound state of the lattice dimer that peels off from the continuum as the strength of atom-atom interactions is increased.  The designations such as ``continuum'' are obviously only qualitative in the case of a finite and possibly even a small number of lattice sites. In the limit of a large number of sites, the the unscaled characteristic frequency of the bound state is
\beq
\frac{E_b}{\hbar} = {\rm sgn}(U)\sqrt{\Omega^2 + U^2}\,.
\label{BSTEN}
\eeq

As has also been noted many times before, the center-of-mass motion encompassed into the lattice momentum $P$ does not completely separate from the internal degree of freedom, the variable $q$. The new feature here is that the rotation phase similarly, indirectly, influences the internal structure of the lattice dimer. In view of Eq.~\eq{OMEGADEF}, for two identical bosons the roles of center-of-mass lattice momentum and rotation phase may be interchanged as it comes to energetics. Any variation with respect to one may just as well be realized by varying the other. 

For attractive atom-atom interactions the bound state is the lowest-energy state, and by our statistical-mechanics assumption it gets prepared at zero temperature. Given the rotation phase $\phi$, the lowest energy occurs for $P \simeq 2\phi$, so that thermal equilibration at low temperature and with fixed rotation phase $\phi$ will in general prepare a finite flow velocity for the molecule.

Also, in the presence of atom-atom interactions the bound state is always separated from the continuum by a nonzero amount, so that adiabatic manipulations of the bound state are possible. One could prepare thermally a stationary  lattice dimer ($P=0$) and then add adiabatically an arbitrary phase $\phi$.  As far as the energetics of the dimer is concerned, this is physically equivalent to generating the center-of-mass lattice momentum $P= -2\phi$. This applies to the spatial structure of the dimer.  The rms size of the bound dimer is $\Delta n= \sqrt2 |\Omega/U|^2$~\cite{JAV10}, so that the dimer could be shrunk to a single site by choosing a rotation speed such that $\Omega=0$. Similarly, the spectroscopy of the dimer~\cite{JAV10} may be controlled.

The final item to understand is measurements of lattice momentum. As we have noted already, the label $q$ even in the MTRF directly corresponds to a lattice momentum measured in the LF. Completely analogously to the earlier analysis of the stationary lattice~\cite{JAV10}, the probability to find a lattice momentum $q$ is proportional to $|A(\half P - q)|^2$. The rotation phase does not directly enter this expression. In the limit $N\gg1$ the unit-normalized (in the sense of the integral over $q$) probability density for the lattice momentum $q$ is found to be
\bea
&&\!\!\!\!\!\!\!\!f(q) = \frac{|{\cal K}(P)|^3}{2\pi\sqrt{1+{\cal K}(P)^2}
}\nonumber\\
&&\!\!\!\!\!\times \frac{1}{\left\{
\cos(q-\half P) +{\rm sgn}[{\cal K}(P)]\sqrt{1+{\cal K}(P)^{\!2}}
\right\}^2}\,,
\label{CPMD}
\eea
with
\beq
{\cal K}(P) = \frac{U}{2J\cos(\half P-\phi)}\,.
\label{NKDEF}
\eeq

Consider past experiments in which repulsively bound lattice dimers were produced with $P\simeq0$ and the lattice momenta of the atoms were subsequently measured. The lattice momenta were predominantly found at the edges of the first Brillouin zone, $q\simeq\pi$~\cite{WIN06}. Suppose now that, after the repulsively bound pairs have been created, it would be possible to spin up the lattice adiabatically to the rotation phase $\phi=\pi$. By Eqs.~\eq{CPMD} and~\eq{NKDEF}, the effect is the same as reversing the sign of the atom-atom interaction. Correspondingly, the lattice momenta would then be found predominantly at the center of the Brillouin zone. This should be contrasted to the observation that adiabatic variation of the rotation speed has no effect on the lattice momentum of a single atom.

A repulsively bound state ($U>0$) lies above the continuum band and an attractively bound state ($U<0$) below. We doubt if any adiabatic method exists that converts one to the other, as in the process the state should move intact across the dissociation continuum. In our thought experiment, according to Eq.~\eq{BSTEN}, the repulsively bound state remains the highest-energy state the entire time while the lattice is set in motion. However, its {\em signature\/} changes to that of an attractively bound state. This is an indirect effect of the energetics. It is as if the rotation effectively changed the sign of the tunneling matrix elements, and here it is the relative sign of the tunneling matrix element and the atom-atom interaction strength that counts.

Still further opportunities for control open up if the atoms are not identical~\cite{PII08,ODO11}, and in particular, if their masses are different. This implies that for the same rotation speed the rotation phases of the two atoms are different. We have discussed  the Hamiltonian and the results that emerge for dissimilar species in detail before for a nonrotating lattice~\cite{ODO11}. Here we briefly comment on the effects of the rotation phases, call them $\phi_1$ and $\phi_2$ for the two species 1 and 2.  These could be bosons, fermions, or one of each.

This time around the expansion coefficient $A(q)$ describing atom 1 with lattice momentum $\half P + q$ and atom 2 with lattice momentum $\half P - q$ is no longer constrained to be an even function of $q$, and the correct form of the inner product now is as one might expect in the first place,
\beq
(\psi_A,\psi_B) = \sum_q A^*(q) B(q)\,.
\eeq
With the definitions
\bea
\Omega&=&\sqrt{J_1^2 + J_2^2 + 2 J_1J_2 \cos(P-\phi_1-\phi_2)}\,,\nonumber\\
\beta &=& \arctan\left[\frac{J_1-J_2}{J_1+J_2}\tan\left(\frac{P-\phi_1-\phi_2}{2}\right)\right] - \half(\phi_1-\phi_2),\nonumber\\
{\cal K}&=&\frac{U_{12}}{2\Omega},\quad \omega=\frac{E}{\hbar\Omega}
\eea
the Schr\"odinger equation becomes
\beq
-\cos(q+\beta) A(q) + \frac{\cal K}{N}\sum_{q'} A(q') = \omega A(q)\,.
\eeq
The branch of the explicit arctan function must be chosen judiciously so that the expression of $\beta$ is a continuous function of its variables. $U_{12}$ is the strength of the interspecies interaction. The eigenvalue equation and its solutions may eventually be written
\bea
&&\frac{1}{N} \sum_q \frac{1}{\omega + \cos(q+\beta)} = \frac{1}{\cal K}\,,\label{DIFEN}\\
&&A(\omega,q) \propto \frac{1}{\omega+\cos(q+\beta)}\label{DIFWF}\,.
\eea

A comparison with the case of two identical bosons first shows that, completely analogously, the sum of the rotation phases of the two atoms always gets subtracted from the center-of-mass lattice momentum.  In the control of the system, $\phi_1+\phi_2$ and $P$ are for the most part equally good knobs to turn.

Next consider the eigenvalue problem~\eq{DIFEN}, \eq{DIFWF}, for the sake of simplicity ignoring the dependence of $\cal K$ on the rotation phases. The values of the parameter $\beta$ then matter only modulo the spacing between the states $q$, or modulo $2\pi/N$. On the other hand, a bound state appears in this system just as for identical bosons and likewise can be modified adiabatically. In such a case an arbitrary value of $\beta$ may have physical relevance.

As an example, let us consider lattice momenta.  Just like in Ref.~\cite{ODO11}, we may find in the limit $N\gg1$ the probability distribution, normalized to unity, that a measurement would find either atom 1 or atom 2 with the lattice momentum $q$,
\bea
&&f_{1,2}(q) = \frac{|{\cal K}(P)|^3}{2\pi\sqrt{1+{\cal K}(P)^2}}\nonumber\\
&&\quad\times \frac{1}{\{{\rm sgn}[{\cal K}(P)]\sqrt{1+{\cal K}(P)^2}+\cos[q-\half P \pm\beta]\}^2};\nonumber\\
&& {\cal K}(P) = \frac{U_{1,2}}{2J\cos[\half(P-\phi_1-\phi_2)]}\,.
\eea
To limit the scope of the present exercise, we assume that the tunneling matrix elements are the same whereupon we have $\beta=\half(\phi_1-\phi_2)$, that it is possible to hold $P+\phi_1+\phi_2\equiv0$ while varying $\phi_1$ and $\phi_2$, and that $\cal K$ is negative. Practicalities aside, this is not as far-fetched as it might sound, since the variable $\phi_{1,2}$ are only defined modulo $2\pi$ and their sum and difference may in general be varied independently even if they are proportional to one another. The species 1 and 2 then predominantly come out with the lattice momenta $-\phi_1$ and $-\phi_2$. The novelty here is that until now we have not produced an example in which the position of the maximum of the lattice momentum distribution is somewhere else than at 0 or $\pm\pi$.

\section{Concluding remarks}\label{CONCLUSION}
We have developed a Hubbard model for atoms in a rotating ring lattice carefully and in great detail in order to correctly identify the physical observables. It turns out that thermal preparation and adiabatic variation of the rotation speed can be used to control the states of both a single atom and a dimer of two atoms, occasionally with unexpected results.

Our Hubbard model could be a rather cursory approximation under many experimental circumstances. For instance, it might happen that the motion of the atoms is not strictly confined to one spatial dimension, or a high rotation speed or atom-atom interactions render the tight-binding approximation questionable. However, the phenomena we have described arise from general principles such as the difficulty of abruptly altering the phase winding of a quantum state around a ring, and should survive modest experimental imperfections.
\bibliography{RotatingLattice}
\end{document}